# IMPLEMENTING CHANGE IN A COMPLEX WORLD

**Dirk Helbing, Computational Social Science @ ETH Zurich, Switzerland**

**Responding to complexity in socio-economic systems: How to build a smart and resilient society?**

The world is changing at an ever-increasing pace. And it has changed in a much more fundamental way than one would think, primarily because it has become more connected and interdependent than in our entire history. Every new product, every new invention can be combined with those that existed before, thereby creating an explosion of complexity: structural complexity, dynamic complexity, functional complexity, and algorithmic complexity. How to respond to this challenge? And what are the costs?

The exponential increase in cybercrime is certainly just one of the undesirable side effects. It now causes damages of the order of 3 trillion dollars each year. The financial crisis is another example. Its damage is estimated to amount to approximately 14 trillion dollars. The increase in the level of global terrorism and international conflict is another problem we must pay attention to. There are further issues related with globalization, such as climate change and international migration. The vulnerability of energy supply and critical infrastructures (e.g. by means of cyber warfare) produces further headaches, and global pandemics, too.

Many of these problems are caused by systemic instabilities, which lead to outcomes that individual actors usually cannot control, despite large amounts of data, advanced technology, and best efforts to keep everything under control. The failure of control typically results from cascade effects, where an incidental anomalous behavior of a system component triggers anomalous behaviors of other system components and so on. Depending on the details of the underlying dynamics, the resulting damage may grow linearly or exponentially. In particular cases, the damage may be even unbounded [1].

Many of humanity's unsolved problems result from such cascading failures. This provides a new perspective on problems ranging from traffic jams over crowd disasters to financial collapse and the spread of crime, terrorism, diseases, conflict and war. As a consequence, understanding the nature of these problems opens up opportunities for new cures.

Recently, many experts have started to hope that Big Data would help us fix the above problems, among others. The idea is that more data would provide more insights, and that knowledge could be turned into power, thereby allowing one to fix the world's problems, perhaps even using an artificial "superintelligence" based on deep learning. So, if one would measure everything and had access to all the data produced on our globe and massive computer power too, could one optimize the course of the world in real time? Could one rule the world like a wise king? This now sounds like a fascinating and plausible perspective [2].

While decision-making was often lacking enough information in the past, Big Data is now offering interesting new perspectives to manage and improve systems. However, there are undesired side effects such as potential discrimination [3] as well as the violation of privacy and undermining of trust [4]. In addition, more data does not necessarily imply better decisions, as demonstrated by the well-known problems of over-fitting (fitting to irrelevant features) and of spurious correlations (identification of patterns that are meaningless). Furthermore, when trying to separate good from bad risks, classification errors are frequent. In other words, no matter how much data are available, mistakes will be made. But if wrongly used, a powerful tool can be very destructive, particularly if one takes large-scale rather than minimally invasive measures. Some of the international wars in the past years, which did not have the intended results, may serve as examples.

Systemic complexity causes additional problems. Complex dynamical systems may be so sensitive to details that it may not be possible to predict their behavior well or even just to calibrate their parameters, which relates to phenomena such as "sensitivity" and "chaos". Moreover, algorithmic complexity may prevent an optimization (or even a proper system analysis) in real time, i.e. even the biggest supercomputers of the world may be too slow (and will probably always be). Finally, as we go on networking the world, systemic complexity increases even faster than data volumes and much faster than processing power. Consequently, the controllability with centralized control approaches will decrease over time! [5]

Therefore, the crucial question is, how to respond to the complexity challenge? How to build resilient systems that are not prone to undesired cascade effects, but recover quickly and well from disruptions? This is primarily a matter of systems design and management. *Modularization* is a well-known principle to make the complexity of a system manageable. This basically means that the organization of a system is broken down into substructures or "units", between which there is a lower level of connectivity or interaction as compared to the connectivity or interaction within the units. This allows one to reduce the complexity within substructures to a manageable level. Furthermore, it decreases interaction effects between units and, with this, undesirable cascade effects.

In principle, of course, the modular units of a system can be organized in a hierarchical way. This can be efficient, when the units (and the interactions between them, including information flows and chains of command) work reliably, with very few errors. However, as much as hierarchical structures help to define accountability and to generate power, control might already be lost if a single node or link in the hierarchy is dysfunctional. This problem can be mitigated by redundancies and *decentralization*. In particular, if the dynamics of a system is hard to predict, local *autonomy* can improve proper adaptation, as it is needed to produce solutions that fit local needs well. More autonomy, of course, requires the decision-makers to take more responsibility, which calls for high-level education and suitable tools, in particular good information systems.

A further important principle that can often support resilience is *diversity*. The benefits of diversity are multifold. First of all, diversity makes it more likely that some units stay functional when the system is disrupted, and that solutions for many kinds of problems already exist somewhere in the system when needed. Second, diversity supports collective intelligence. Third, the innovation rate typically grows with diversity, too. However, diversity also poses challenges, as we know, for example, in intercultural interactions. For this reason, interoperability is important. I will come back to this issue below.

Finally, how can one control a complex dynamical system in a distributed way? This can be done using the principle of *(guided) self-organization* [6,7]. In complex systems, where many system components respond to each other in non-linear ways, the outcome is often the emergence of macro-level structures, properties, and functions. The kind of outcome depends, of course, on the details of these interactions. But modifying the interactions allows one to let other structures, properties, and functions emerge. The disciplines needed to find the right kinds of interactions to obtain a desirable outcome are called "complexity science" and "mechanism design".

Even with simple local interactions, it is possible to generate a surprisingly rich spectrum of often complex structures, properties and functions. One particularly favorable feature of self-organization is that the resulting structures, properties and functions occur by themselves and very efficiently, by using the forces within the system rather than forcing the system to behave in a way that is against "its nature". Moreover, the so resulting structures, properties and functions are stable with regard to moderate perturbations, i.e. they tend to be resilient against disruptions, as they would tend to reconfigure themselves according to "their nature".

But how to determine suitable interaction rules to let a system produce a certain desired outcome? There are different possibilities. Computer simulations allow one to study the self-organization of complex dynamical systems in a computer, if the interactions are simple enough and well enough defined. Otherwise, to get an idea what outcomes the interactions of real human beings might produce, one can perform lab experiment or web experiments using Amazon Mechanical Turk. Furthermore, interactive online games have become a tool for the exploration of socio-economic interactions. Finally, it will be worth identifying the mechanisms on which the cultures of the world are based. These cultural mechanisms, in fact, are of high importance for well-functioning societies and their resilience to disruptions. Surprisingly, most of these mechanisms are not explicitly known, but are "internalized" subconsciously. If they were known, however, we could combine the many success mechanisms of the world's cultures in new ways.

Interactions produce *"externalities"*, i.e. external effects, but these can usually be changed by introducing or modifying *feedback loops* in the system. Such feedbacks allow the system components to adapt to the local conditions in ways that restore the normal functionality. In economic systems, feedback mechanisms are often produced by financial costs or rewards, while in social systems it is common to use incentives and sanctions [8]. However, certain kinds of information exchange and coordination mechanisms are even more efficient ("altruistic signaling", for instance) [7]. It is also important to consider that one kind of feedback mechanism (such as money) is usually too restricted to let a complex socio-economic system self-organize, and therefore a multi-dimensional value exchange system is needed, as I have recently proposed it [9].

In fact, many chemicals or pharmaceutical drugs cannot be produced by controlling a single variable such as the concentration of a particular ingredient. Instead, one needs to control the temperature, pressure, and concentrations of many ingredients. In a similar way, our body will not do well, if we increase the quantity of just one substance, e.g. the amount of water we drink. We need to have enough carbon hydrates, proteins, vitamins, and minerals as well. Therefore, to create a better working economy, the establishment of a multi-dimensional value exchange system is inevitable.

The multi-dimensional value exchange system would be best built on the externalities that matter, i.e. all the in- and outputs. Desirable outputs would be represented by positive numbers ("gains") in a specific dimension related to that particular kind of output, and undesirable ones by negative numbers ("losses"). Desirable inputs would be represented by negative numbers ("costs"), and undesirable inputs should be avoided. In other words, to enable a self-organizing economy, externalities must be measured in real-time to allow for real-time feedbacks, and those feedbacks would be created by the multi-dimensional value exchange system I propose.

Interestingly, the real-time measurement of externalities becomes increasingly possible now, thanks to the spread of the "Internet of Things", i.e. of networks of sensors that can communicate with each other in a wireless way. For this purpose, my collaborators and I have recently proposed to build a participatory information platform as a Citizen Web, which we call the "Planetary Nervous System" [10]. With this enabling technology, we can finally make the "invisible hand" work. That is, 300 years after its invention, we can perform the measurement of externalities and feed them back on the decision-making entities (people, institutions, companies, or even algorithms) in such a way that efficient AND desirable outcomes are produced. For example, one can build assistant systems to dissolve traffic jams or produce fluent traffic flows in cities. One could also build an assistant system to stabilize global supply chains and thereby reduce the bullwhip effects that would otherwise produce booms and recessions. Furthermore, one could build digital assistant systems to support cooperation and avoid conflict [6]. These "Social Technologies" would help one to ensure favorable outcomes of interactions for all sides.

In fact, interactions between two entities (be it people, companies, or institutions) can basically have four possible outcomes: (1) If an interaction would be lossful for both entities, as it is often the case in conflicts and wars, the interaction should be avoided. (2) If the interaction would be favorable for one side, but bad for the other and lossful overall, the interaction should be also avoided, and to ensure this, the second entity must be protected from exploitation by the first one. (3) If the interaction would again be favorable for one side and bad for the other, but positive overall, it can be turned into a win-win situation by means of a value transfer. (4) Finally, if the interaction would be beneficial for both sides, one should engage in it, but one might decide to share the overall benefits in a fairer way by means of a value exchange.

Digital assistants could support us in all these situations. They could help to create situational awareness, including the potential side effects and risks implied by certain decisions and (inter)actions. Without such assistants, we would certainly overlook many opportunities for beneficial interactions we could actually engage in. Digital assistants could also help us to organize protection against exploitation, which would otherwise deteriorate the overall state of the system. And finally, Social Technologies could support us with multi-dimensional value exchange, as I discussed it before. Social Technologies can assist us particularly in avoiding the systemic instabilities that I discussed in the beginning of this chapter as the main source of our unsolved problems. This might also include digital assistance to reduce "tragedies of the commons" such as environmental exploitation, overfishing, or global climate change.

In summary, instabilities in complex systems and the often resulting large-scale cascading failures are the underlying reasons for some of the greatest unsolved problems in the world. They result from wrong system designs and management approaches, which lead to uncontrollable outcomes, despite massive amounts of data, modern technology, and best intentions. However, a paradigm shift in the way we are creating and managing these systems could solve our problems. One would mainly have to engage in a distributed systems approach, characterized by modular designs, distributed control, and self-organization. This also applies to our entire economy [11]. Diversity is another relevant ingredient, which is important for resilience, innovation, and collective intelligence. However, in the past we have often had difficulties to handle diversity. Digital assistants can support us in this, such that we will become increasingly able to reap the benefits of diversity, which is also a key factor of economic success and social well-being.

Finally, we have not made sufficient use of the success principles underlying the diverse cultures in the world. This can now be changed. But to make various systems interoperable and to produce favorable outcomes of interactions, one needs to measure the diverse externalities, and feed them back by means of a multi-dimensional value exchange system. The "Planetary Nervous System" is the enabling technology for this. Combined with the insights of complexity science, this will finally allow us to let the "invisible hand" work for us, creating a self-organization of complex dynamical systems that produces the systemic structures, properties and functions we want [12].

### Further Reading